\documentclass[aps,prl,twocolumn,groupedaddress,groupedaddress]{revtex4}  
\usepackage{amsfonts,ulem,bbold}
\usepackage{amssymb,amsmath}%,graphicx,wrapfig,epsfig}
\usepackage[usenames]{color} 

\newcommand{\be}{\begin{equation}}  
\newcommand{\ee}{\end{equation}}  
\newcommand{\bea}{\begin{eqnarray}}  
\newcommand{\eea}{\end{eqnarray}}  
\newcommand{\p}{\partial}  
\newcommand{\nn}{\nonumber}  
  
\newcommand{\mA}{\mathbb A}
  
\newcommand{\mB}{\mathbb B}

\newcommand{\mK}{\mathbb K}  
\newcommand{\mX}{\mathbb X}  
\newcommand{\one}{\mathbb 1} 
\newcommand{\I}{{\bf I}}
\newcommand{\invI}{{\bf I}^{-1}}

\DeclareMathOperator{\tr}{tr}

\begin{document}   

\title{Resolving the Ghost Problem in non-Linear Massive Gravity} 

\author{S. F. Hassan}
\author{Rachel A. Rosen}
\affiliation{Department of Physics \& The Oskar Klein Centre,
  Stockholm University, AlbaNova University Centre, SE-106 91
  Stockholm, Sweden}

\begin{abstract}
We analyze the ghost issue in the recently proposed models of
non-linear massive gravity in the ADM formalism. We show that, in the
entire two-parameter family of actions, the Hamiltonian constraint is
maintained at the complete non-linear level and we argue
for the existence of a non-trivial secondary constraint. This implies
the absence of the pathological Boulware-Deser ghost to all orders. To
our knowledge, this is the first demonstration of the existence of a
consistent theory of massive gravity at the complete
non-linear level, in four dimensions.
\end{abstract}  
  
\maketitle
  
{\bf Introduction and Summary:} 
The search for a consistent theory of massive gravity is motivated by
both theoretical and observational considerations. Since the
construction of a linear theory of massive gravity by Fierz and Pauli
in 1939 \cite{FP1,FP2}, a proof of the existence of a consistent
non-linear generalization has remained elusive, making it a
theoretically intriguing problem. On the observational side, the
recent discovery of dark energy and the associated cosmological
constant problem has prompted investigations into long distance
modifications of general relativity. An obvious such modification is
massive gravity.

Theories of massive gravity generically suffer from a ghost
instability. The origin of this problem can be understood as follows.
In general relativity the four constraint equations of the theory
along with four general coordinate transformations remove four of the
six propagating modes of the metric, where a propagating mode refers
to a pair of conjugate variables. The total number of propagating
modes is thereby reduced to the physical two modes of the massless
graviton.  In contrast, in massive gravity the four constraint
equations generically remove the four non-propagating components of
the metric while the general covariance is broken. Thus the theory
will contain six propagating modes of which only five correspond to
the physical polarizations of the massive graviton. The remaining mode
is a ghost.

The question then is whether it is possible to construct a theory of
massive gravity in which one of the constraint equations and an
associated secondary constraint eliminate the propagating ghost mode
instead. The linear Fierz-Pauli theory succeeds in eliminating the
ghost in this way. But Boulware and Deser \cite{BD} showed that the
ghost generically reappears at the non-linear level. More recently,
progress was made in \cite{AGS} by observing that the ghost is related
to the longitudinal mode of the Goldstone bosons associated with the
broken general covariance. This greatly simplifies the analysis of the
ghost problem in the so-called decoupling limit which isolates
non-linear effects in the ghost sector. Based on this approach a
procedure was outlined in \cite{AGS,CNPT} to avoid the ghost
order-by-order by tuning the coefficients in an expansion of the mass
term in powers of the metric perturbation and of the Goldstone mode.
In 2010 de Rham and Gabadadze \cite{dRG2} successfully obtained such
an expansion which is ghost free in the decoupling limit. Later in
\cite{dRGT}, these perturbative actions were resummed into fully
non-linear actions resulting in a two-parameter family of theories.
This was the first successful construction of potentially ghost free
non-linear actions of massive gravity. Also in \cite{dRGT}, one of
these resummed actions was analyzed in the ADM formalism \cite{ADM}
and it was argued to be ghost free to fourth order in metric
perturbations around flat space. In \cite{ACM} it is claimed that the
ghost still appears at the fourth order. (For a review of recent
developments in massive gravity, see \cite{Hinterbichler}.) The
present work addresses the ghost issue at the non-perturbative level.

The systematics and generality of these potentially ghost free actions
are studied in \cite{HR}. In particular, they are presented as a
two-parameter generalization of a minimal extension of the Fierz-Pauli
theory. In this work we show that the entire two-parameter family of
actions is ghost free at the full non-linear level. Starting with the
minimal theory in the ADM formalism, we show that the lapse $N$ is
indeed a Lagrange multiplier leading to a Hamiltonian constraint on
the propagating modes. We also show that
  the same analysis extends to the full two-parameter generalization
  of the minimal theory. We then argue that this Hamiltonian constraint
gives rise to a secondary constraint.  These are enough to eliminate a
single propagating mode, ensuring that the theory contains only five
propagating degrees of freedom appropriate for the spin-2 massive
graviton. Thus the Boulware-Deser ghost is eliminated.
\vskip.1cm

{\bf Non-Linear Massive Gravity:} 
In the Fierz-Pauli theory, linearized general relativity in flat space
is extended by the addition of a mass term for the metric fluctuations
$h_{\mu \nu} = g_{\mu \nu}-\eta_{\mu \nu}$,
\be
\label{FP}
\frac{m^2}{4}(h_{\mu \nu} h^{\mu \nu}-h^{\mu}_{\mu} h^{\nu}_{\nu}) \, .
\ee
To construct non-linear generalizations of the Fierz-Pauli mass term,
an additional non-dynamical metric $f_{\mu \nu}$ is invariably
required. In the recently developed potentially ghost free theories,
the basic building block is a matrix of the form $\sqrt{g^{-1}f}$
\cite{dRGT,HR}, where the square root of the matrix is defined 
such that $\sqrt{g^{-1}f} \sqrt{g^{-1}f} = g^{\mu \lambda}f_{\lambda
  \nu}$. In particular, the minimal extension of the Fierz-Pauli
action with zero cosmological constant is given by \cite{HR},
\be
\label{actmin}  
S=M_p^2\int d^4x\sqrt{-g}\,\left[R -2m^2\,(\tr \sqrt{g^{-1}f}-3)\right]\, .
\ee  
Our ghost analysis is based on this action for the case of a flat
$f_{\mu \nu}$ so that in the physical gauge
$f_{\mu\nu}=\eta_{\mu\nu}$. Recent studies of massive gravity have
primarily focused on this case \cite{dRG2,dRGT}. 

However, with this choice of $f_{\mu \nu}$, the minimal action does
not have a Vainshtein mechanism \cite{V} and thus exhibits the vDVZ
discontinuity \cite{vDVZ1,vDVZ2}, as shown in \cite{Koyama}. To be
compatible with observations one must consider theories with
additional higher order interactions that could induce a Vainshtein
mechanism (see, e.g., \cite{TMN,Koyama,CP}). We will show that our
analysis naturally extends to the entire family of such 
  actions without any modification. 

The most general non-linear massive gravity theories that are
potentially ghost-free are given by a two-parameter family of actions.
Defining a matrix $\mK$ so that $\sqrt{g^{-1}f}=\one+\mK$, these
actions can be written as \cite{dRGT},
\be
\label{act}  
S=M_p^2\int d^4x\sqrt{-g}\,\bigg[R
 + 2 m^2\,\sum_{n=2}^4\alpha_n  e_n(\mK)\bigg]\,,  
\ee
with $\alpha_2=1$ and where the $e_k$ are defined in (\ref{ek})
below. 
 
For our purposes it will be easier to work with an equivalent
formulation of (\ref{act}) \cite{HR},
\be
\label{act2}  
S=M_p^2\int d^4x\sqrt{-g}\,\bigg[R +2m^2 \sum_{n=0}^{3} \beta_n\,
  e_n(\sqrt{g^{-1} f})\bigg] ,
\ee 
where the $\beta_n$ are given in terms of the $\alpha_n$ of
(\ref{act}) as,  
\bea 
&\beta_0 = 6-4\alpha_3+\alpha_4 \, , ~~~&\beta_1 =
-3+3\alpha_3-\alpha_4 \, ,\nonumber \\  
&\beta_2 = 1-2\alpha_3+\alpha_4 \,, ~~~&\beta_3 =\alpha_3-\alpha_4\,.   
\eea
The $e_k(\mX)$ are elementary symmetric polynomials of the eigenvalues
of $\mX$.  For a generic $4\times 4$ matrix they are given by, 
\begin{align}
\label{ek}   
e_0(\mX)&= 1  \, , \hspace{.3cm}
e_1(\mX)= [\mX]  \, ,\hspace{.3cm} 
e_2(\mX)= \tfrac{1}{2}([\mX]^2-[\mX^2])\,\nn \\  
e_3(\mX)&= \tfrac{1}{6}([\mX]^3-3[\mX][\mX^2]+2[\mX^3])
\, ,\nonumber \\   
e_4(\mX)&=\tfrac{1}{24}([\mX]^4-6[\mX]^2[\mX^2]+3[\mX^2]^2   
+8[\mX][\mX^3]-6[\mX^4])\, ,\nonumber \\  
e_k(\mX)&= 0 ~~{\rm for}~ k>4 \, ,
\end{align} 
where the square brackets denote the trace. The action (\ref{act2})
contains terms that are at most third order in $\sqrt{g^{-1} f}$
rather than fourth order as in (\ref{act}). When $\alpha_3 =
\alpha_4=0$ one obtains the resummed theory for which the ghost
analysis was performed in \cite{dRGT} to fourth order. When $\alpha_3
= \alpha_4 =1$ one obtains the minimal action (\ref{actmin}). After
treating the minimal action, we will extend the ghost analysis to the
most general case (\ref{act2}), for arbitrary $\beta_n$.

\vskip.1cm

{\bf The Hamiltonian constraint:}
Let us recapitulate the counting of degrees of freedom in standard
massless general relativity. In the ADM formulation \cite{ADM}, the ten 
components of the metric are parametrized as 
\be 
N =(-g^{00})^{-1/2}\,, \qquad N_i = g_{0i}\,, \qquad \gamma_{ij}=
g_{ij}\,.  
\ee 
The $\gamma_{ij}$ describe six potentially propagating modes.  The
action written in terms of canonical variables is linear in the
non-propagating modes $N$ and $N_i$ (collectively, $N_\mu$). Thus
the $N_\mu$ equations of motion are constraints on the $\gamma_{ij}$
and their conjugate momenta $\pi^{ij}$.  Along with the general
coordinate transformations they eliminate four out of six propagating
modes, a propagating mode referring to a pair of conjugate
variables. The $N_\mu$ are determined by the remaining equations, thus
leaving two propagating modes corresponding to a spin-2 graviton.

In a generic non-linear extension of massive gravity, the mass term
depends non-linearly (but still algebraically) on the $N_\mu$.  The
corresponding equations of motion determine these non-dynamical
variables in terms of $\gamma_{ij}$ and $\pi^{ij}$, keeping all six of
the propagating modes undetermined.  Five propagating modes of
$\gamma_{ij}$ correspond to the massive graviton, while the sixth one
is a ghost, called the Boulware-Deser mode \cite{BD}.  A ghost free
theory of massive gravity must maintain a single constraint on
$\gamma_{ij}$ and $\pi^{ij}$ along with an associated secondary
constraint to eliminate this ghost-like sixth mode.  Below we show
that this is indeed the case for the non-linear massive gravity
actions described above.

Let us first consider the minimal massive gravity action
(\ref{actmin}). In the ADM parameterization the Lagrangian ${\cal L}$
is given by, 
\be
\pi^{ij}\partial_t \gamma_{ij} + N R^0 +N^i
R_i-2m^2\sqrt{\gamma}\, N\,\left(\tr \sqrt{g^{-1}\eta}-3\right) ,
\ee
where (with $N^i=\gamma^{ij}N_j$),
\bea
\label{ADM}
(g^{-1}\eta)^\mu_{\,\,\nu} =\frac{1}{N^2} 
\left( \begin{array}{ccc}
1 &~~& N^l \delta_{lj} \\
-N^i && (N^2 \gamma^{il}-N^iN^l)\delta_{lj} 
\end{array} \right) \, .
\eea
Here and in what follows we use $\sqrt{\gamma}$ to denote
$\sqrt{\det\gamma_{ij}}$. 

The action (\ref{ADM}) is highly nonlinear in $N_\mu$ and thus it
might appear that there are no constraint equations for the
propagating degrees of freedom.  However, if the four $N_\mu$
equations only depend on three combinations of $N$ and
$N_i$, the fourth equation can be used to determine the sixth mode of
$\gamma_{ij}$ in terms of remaining modes.   

To show that is the case, we start by assuming that three such
combinations $n^i$ exist.  Then, after writing $N^i$ in terms of
$n^i$, the massive gravity actions should satisfy the following two
criteria,
\begin{enumerate}
\item The action is linear in $N$ so that the $N$ equation of motion
  becomes a constraint on the other fields.
\item The equations of motion for the $n^i$ are independent of $N$ and
  hence are algebraically solvable for the $n^i$.
\end{enumerate}
Thus the $N$ equation becomes a constraint on the $\gamma_{ij}$ and
$\pi^{ij}$.  Along with a secondary constraint, this removes the
ghost.  $N$ itself is non-dynamical and is expected to be determined
by the remaining equations, as in GR \cite{ADM}.  We will see that
when criterion 1 is satisfied, 2 will follow automatically.

Criterion 1 means that the change of variables must be linear in $N$,
hence we consider, 
\be
\label{L}
N^i = (\delta^i_j + N D^i_{\,\,j}) n^j \, .
\ee
The matrix $D \equiv D^i_j$ is determined by requiring that the mass
term is linear in $N$. Indeed that will be the case if the square-root
matrix has the form, 
\be
N \sqrt{g^{-1}\eta} = \mA +N \mB \, ,
\ee
where matrices $\mA$ and $\mB$ are independent of $N$. Then,
\be
g^{-1}\eta=\frac{1}{N^2} \mA^2 +\frac{1}{N} (\mA \mB+\mB \mA)+\mB^2\,. 
\label{AB}
\ee
On the other hand, to write $g^{-1}\eta$ in terms of the new
variables, let's assemble the $n^i$ into a column vector $n$, with
transpose $n^T$, and write $\eta=diag\{-1,\I \}$, where, 
\be
\I=\delta_{ij}\,,~~  \invI=\delta^{ij}\,,~~  {\rm whereas} ~~
\one=\delta^i_j \, . 
\ee 
Then, writing (\ref{ADM}) in terms of the
variables (\ref{L}) and identifying the resulting expression with
(\ref{AB}), we read off,
\be
\label{mA}
\mA = \frac{1}{\sqrt{1-n^T\I n}} 
\left( \begin{array}{ccc}
1  & ~ & n^T\I \\
-n &    &-nn^T\I 
\end{array} \right) \,,
\ee
\be
\label{mB}
\mB = \left( \begin{array}{ccc}
0 & ~ & 0 \\
0 &   &\sqrt{(\gamma^{-1}-Dnn^TD^T)\I} 
\end{array} \right) \, ,
\ee
and
\be
(\sqrt{1-n^T\I\, n})\, D=\sqrt{(\gamma^{-1}-Dn n^TD^T)\I} \, .
\label{D}
\ee
This last equation can be easily solved for $D^i_j$. However, for the
arguments that follow we only need the equality (\ref{D}) and not the
explicit solution. Note that the transformation (\ref{L}) contains no
time derivatives and can be shown to be invertible.

A crucial property of $D$ is that $D^i_{\,\,l}\delta^{lj}$ is
symmetric, 
\be
D\invI = (D\invI)^T \,.
\label{Dsym}
\ee 
This can be seen from (\ref{D}) combined with the relation
$(\sqrt{M\I})\invI= \invI(\sqrt{\I M})$ which holds for any matrix
$M^{ij}$. 

Now in terms of the new variables $n^i$, the action is linear in $N$, 
meeting the first criterion,
\begin{align}
{\cal L} =& \pi^{ij}\partial_t \gamma_{ij} + N
R^0+R_i (\delta^i_j+ND^i_{\,\,j})n^j \nonumber \\ 
& -2m^2\sqrt{\gamma}\, \left[\sqrt{1-n^T \I n}\right.\nonumber \\
& \left. +
N\tr(\sqrt{\gamma^{-1}\I - Dnn^TD^{T}\I})   -3N\right]\, .  
\label{LADM}
\end{align}
The symmetry property of $D$ along with expression (\ref{D}) can now
be used to show that the $n^i$ equations of motion are independent of
$N$ as demanded by criterion 2. Indeed, using
$\delta\tr\sqrt{M}=\tfrac{1}{2} \tr(\sqrt{M}^{\,-1}\delta M)$ to
differentiate the trace term, one gets, after some manipulations, the
$n^k$ equation of motion,
\be
\left(R_i+\frac{2m^2\sqrt{\gamma}\,n^l\delta_{li}}{\sqrt{1-n^r
    \delta_{rs} n^s}} \right)
\left[\delta^i_k+N\frac{\p}{\p n^k}\left(D^i_{\,j}n^j\right)\right]
=0\,. 
\nn
\ee
The expression in the square brackets is the Jacobian of the
transformation (\ref{L}) and is non-zero. Hence the $n^i$ equations
are, 
\be
(\sqrt{1-n^r \delta_{rs} n^s})\,R_i+2 m^2\sqrt{\gamma}\,
n^l\delta_{li}=0 \, .
\label{neom-final}
\ee
These can be readily solved to determine $n^i$ in terms of
$\gamma_{ij}$ and the conjugate momenta $\pi^{ij}$,
\be
n^i=- R_j \delta^{ji}\left[4m^4\det{\gamma} +R_k \delta^{kl} 
R_l \right]^{-1/2}\, .
\ee
This solution implies that $\sqrt{1-n^T \I n}$ is real.

The $N$ equation of motion is, 
\be
R^0+R_i D^i_{\,j}n^j -2m^2\sqrt{\gamma}\left[
\sqrt{1-n^r \delta_{rs} n^s}\,\, D^k_{~k} -3 \right]=0 \, .
\label{Neom}
\ee
Using the $n^i$ solution, this clearly becomes a constraint on the 12
components of $\gamma_{ij}$ and $\pi^{ij}$. Note that in the limit
that $m^2 \rightarrow 0$, (\ref{neom-final}) and (\ref{Neom}) reduce
to the four constraints of general relativity.

\vskip.1cm

{\bf The general action:}
We now extend the analysis of the previous section to the full
two-parameter generalization of the minimal theory. First 
consider the next higher term in $\sqrt{g^{-1}\eta}$ in the action
(\ref{act2}), given by,
\be
\label{e2}
e_2(\sqrt{g^{-1}\eta}) = \frac{1}{2} \left[(\tr \sqrt{g^{-1}\eta})^2 
-\tr g^{-1}\eta \right] \, .
\ee
To express this in terms of the variables defined in the previous
section, note that the matrix $\mA$ has the property $\tr (\mA^k) =
(\tr \mA)^k$. The potential (\ref{e2}) then gives, 
\be
N\,e_2=
\frac{1}{2} \left[2\right(\tr \mA \tr \mB-\tr \mA \mB
   \left)+N\left((\tr \mB)^2 -\tr \mB^2\right) \right] \,.
\ee
This is linear in the lapse $N$ and thus also satisfies our first
criterion.  Varying with respect to $n^k$ gives, 
\begin{multline}
\frac{\delta}{\delta n^k} \left(N\,e_2\right)= \\ 
\hfill -\left(n^l \delta_{li}D^m_{~m}-n^l \delta_{lm}D^{m}_{~i} 
\right) \left[\delta^i_k+N\frac{\p}{\p n^k}\left(D^i_{\,j}n^j\right) 
\right]. 
\end{multline}
It is straightforward to show that the next term in the
potential, $N \,e_3(\sqrt{g^{-1}\eta})$, is also linear
in the lapse $N$, and, through a more involved analysis, determine the
corresponding contribution to the $n^k$ equation of motion. 

Combining these results with those from the previous sections, the
complete equations of motion for $n^i$ are, 
\bea
&&R_i-2 m^2\sqrt{\gamma}\bigg(\beta_1 \frac{n^l\delta_{li}}{\sqrt{1-n^r
\delta_{rs}n^s}}+\beta_2 n^l\Big[\delta_{li}D^k_{~k}-\delta_{lk}D^{k}_{~i}
\Big] \nonumber
\\ 
&&+\beta_3 (\sqrt{1-n^r \delta_{rs} n^s}) \, n^l\delta_{lk} 
\Big[D^k_{~m}D^m_{~i}-D^k_{~i}D^m_{~m} \qquad\quad  \nonumber \\
&&+\tfrac{1}{2}D^m_{~m}D^j_{~j}\delta^k_i-
\tfrac{1}{2}D^m_{~j}D^j_{~m}\delta^k_i
 \Big] \bigg)=0\,.    
\label{neom-general}
\eea
As in the previous sections, these equations are independent of $N$
and can be used to eliminate $n^i$. The $N$ equation
is then the constraint on $\gamma_{ij}$ and $\pi^{ij}$,
\begin{align}
&R^0+R_i D^i_{\,j}n^j  \\
&+2m^2\sqrt{\gamma}\Big[\beta_0+\beta_1 \tr \mB
+ \tfrac{1}{2} \beta_2 \left\{ (\tr \mB)^2 - \tr \mB^2 \right\}
\nonumber \\ 
&+ \tfrac{1}{6} \beta_3 \left\{ (\tr \mB)^3 - 3 \tr \mB \tr \mB^2+2
\tr \mB^3 \right\} \Big]=0 \, . \nonumber 
\label{Neom2}
\end{align}

{\bf The secondary constraint:} 
We now argue that the Hamiltonian constraint gives rise to a secondary
constraint (for a proof, see \cite{HR-const}, completed while this work
was in review). This implies that the 12 dimensional phase space of
the dynamical variables $\gamma_{ij}$ and $\pi^{ij}$ has only 10
degrees of freedom, corresponding to the five polarizations of the
massive graviton.   
  
We have shown that, upon integrating out the shift $N^i$, 
the Lagrangian (\ref{LADM}) remains linear in the lapse $N$,
\be
\label{LC2}
{\cal L}= \pi^{ij}\partial_t \gamma_{ij} - {\cal H}_0
(\gamma_{ij},\pi^{ij})+N\, {\cal C} (\gamma_{ij},\pi^{ij}) \, . 
\ee
A secondary constraint is obtained by demanding that the primary
constraint ${\cal C}$ is independent of time on the constraint surface.
In the Hamiltonian formulation this condition is given in terms of the
Poisson bracket,  $\left\{{\cal C}, \, H \right\} \approx 0$, where
$H=\int d^3x( {\cal H}_0- N{\cal C})$.  If $\{{\cal C}(x),\,{\cal
  C}(y)\} \approx 0$, then this condition is independent of $N$ and
thus becomes a constraint on $\gamma_{ij}$ and $\pi^{ij}$, 
\be
{\cal C}_{_{(2)}} \equiv \left\{{\cal C}, \, H_0 \right\}\approx 0 \, ,
\ee
where now  $H_0=\int d^3x\, {\cal H}_0$.

By construction, the Lagrangian (\ref{LC2}) reproduces the Fierz-Pauli
Lagrangian at lowest order in the fields,  
\be
{\cal H}_0\simeq{\cal H}_0^{FP}+{\cal O}(\gamma^3,\pi^3), ~~~~ {\cal
  C} \simeq{\cal C}^{FP}+{\cal O}(\gamma^2,\pi^2)\,. 
\ee 
Hence one can compute, 
\be
{\cal C}_{_{(2)}} \simeq {\cal C}_{_{(2)}}^{FP} +{\cal O}(\gamma^2,\pi^2)\, ,
\ee
where ${\cal C}_{_{(2)}}^{FP} $ is neither identically zero nor equal
to ${\cal C}^{FP} $. Now, in the Fierz-Pauli case, we know that
$\{{\cal C}^{FP}(x),\,{\cal C}^{FP}(y)\} \approx 0$. Thus at lowest
order in the fields there exists a 
non-trivial secondary constraint.  As long as $\{{\cal C}(x),\,{\cal
  C}(y)\} \approx 0$ continues to hold at the non-linear level (for a
proof, see \cite{HR-const}), then ${\cal C}_{_{(2)}}$ remains a
non-trivial secondary constraint at the non-linear level as well.
Moreover, as can be seen from the Fierz-Pauli structure, enforcing
$\left\{{\cal C}_{_{(2)}}, \, H \right\} \approx 0$ will result in an
equation for $N$, rather than a tertiary constraint.  Thus no further
degrees of freedom are removed in this way. 

\vskip.1cm

{\bf Discussion:} This work demonstrates the existence of non-linear
theories of massive gravity that do not suffer from the Boulware-Deser
ghost instability.  Note that, in order not to violate the
constraints found above, the coupling of the metric to matter must
also be linear in the lapse and shift functions.  The minimal coupling
of General Relativity automatically satisfies this requirement and
hence will not change the arguments presented here.

It should be emphasized that while it is common to discuss the ghost
in terms of St\"uckelberg fields \cite{AGS,CNPT,dRG2,dRGT}, the
Boulware-Deser instability \cite{BD} is, strictly speaking, due to the
loss of the Hamiltonian constraint. We have shown that the massive
actions (\ref{act2}) precisely avoid this problem.

\vskip.1cm

{\it Acknowledgements:} 
We would like to thank F. Berkhahn, C. de Rham,
  S. Folkerts, G. Gabadadze, K. Hinterbichler, S. Hofmann, A. Pritzel,
  A. Schmidt-May, B. Sundborg, A. Tolley and N. Wintergerst for
discussions and comments on the draft. R.A.R is supported by the
Swedish Research Council (VR) through the Oskar Klein Centre.

\end{document}